\documentclass[smallcondensed]{svjour3}
\usepackage[latin2]{inputenc}
\usepackage{graphicx}
\usepackage{amsmath}
\usepackage{blindtext, graphicx}
\usepackage{amsmath}
\usepackage{algorithm}
\usepackage{algorithmic}
\usepackage[numbers,sort&compress]{natbib}
\usepackage{hyperref}

\usepackage{amsmath,amssymb,amsfonts}
\usepackage{algorithmic}

\usepackage{textcomp}
\usepackage{xcolor}
\usepackage{epsfig,subfigure,multirow,amsmath}
\usepackage{mathptmx}
\usepackage{graphicx}
\usepackage{epstopdf}
\usepackage{rotating, booktabs, xcolor, siunitx}
\usepackage{tabularx}
\usepackage{tabulary}
\usepackage{colortbl}
\usepackage{enumitem} 
\usepackage{amsmath}

\begin{document}

%\selectlanguage{english} %%% remove comment delimiter ('%') and select language if required

\title{A Fuzzy Community-Based Recommender System Using PageRank}

\author{
  Maliheh Goliforoushani \and
  Radin Hamidi Rad
 \and
    Maryam Amir Haeri}

\def\makeheadbox{\relax}

\date{}


\institute{Maliheh Goliforoushani \at
              Department of Electrical \& Computer Engineering, Islamic Azad University Tehran North Branch, Tehran, Iran \\            
              \email{maliheh\_goliforoushani@iau-tnb.ac.ir}           %  \\
           \and
           Radin Hamidi Rad \at
              Department of Computer Engineering and Information Technology, Amirkabir University of Technology, Tehran, Iran\\
              \email{radin.h@aut.ac.ir}
              \and
              Maryam Amir~Haeri \at Department of Computer Engineering and Information Technology, Amirkabir University of Technology, Tehran, Iran\\
              \email{haeri@aut.ac.ir}}
\maketitle

\begin{abstract}

Recommendation systems are widely used by different user service providers specially those who have interactions with the large community of users. 
This paper introduces a recommender system based on community detection. The recommendation is provided using the local and global similarities between users. The local information is obtained from communities, and the global ones are based on the ratings. Here, a new fuzzy community detection using the personalized PageRank metaphor is introduced. The fuzzy membership values of the users to the communities are utilized to define a similarity measure.  The method is evaluated by using two well-known datasets:  MovieLens and  FilmTrust. The results show that our method outperforms recent recommender systems. 
\end{abstract}

\section{Introduction}

Nowadays, we are faced with the rapid growth of information on the Internet which is leading to information overload. It is a difficult task for the Internet users to find their desired information among the massive amounts of information on a reasonable time. Therefore, to solve this problem, recommender systems (RS) can be a huge game changer for service providers specially e-commerce services due to the fact that small increase in user interaction percentages can lead to a giant profit gain for the market.

 Recommender systems are software tools and techniques that help users for decision-making process such as what item to buy, what movie to watch, or what music to listen to \cite{ricci2015recommender}. Recommendations are usually personalized and recommender systems are trying to suggest products and services based on user preferences. There are also some non-personalized recommendations which can be generated more easily, typical examples include the top-ten lists of books or movies etc. Although such recommendations may be useful in some cases, however they are not usually considered  by RS research.

 Recommender systems can be divided into two general categories: content-based and collaborative filtering approaches. Collaborative Filtering is the most popular approach and more widely applied in recommender systems since it overcomes some of the limitation of content-based ones. This approach recommends the target user the items other users with similar interests rated and liked in the past. Collaborative Filtering methods can be categorized in the two generic types of neighborhood and model-based methods. One of the primary approaches proposed for function of item recommendation is neighborhood-based recommendation that is the most popular method for this issue. Neighborhood-base collaborative filtering uses user-item ratings which are stored in the system, to predict ratings for new items and this can be accomplished in two ways which are known as user-based or item-based recommendation. When a user asks for a recommendation, user-based approaches evaluate the user interests for all of the unrated items based on rating information from similar users' recorded data. In the same way, the item-based systems compute recommendations by discovering items that are similar to other items the target user has liked \cite{sarwar2001item}.

Generally, the purpose of recommendation systems is to personalize user's experience based on his prior activities and information collected from user profiles, so for generating high quality and reliable recommendations, information such as enhanced user profiles, various types of interaction and relation among users, and some notions like community can be exploited from social networks that have acquired enormous popularity over the past few years. In fact studying social behavior of users can be a very helpful tool to obtaining an accurate vision to user's similarities. Communities are groups of people in social networks who have extensive communications with each other than to the rest of the network. Therefore,  community detection of a social network can identify people who have similar tastes and provide a new hidden set of concepts that cannot be extracted from other regular sources.

Taking all the aforementioned into account, using social computing techniques to achieve better performance in recommendation tasks seems reasonable. In this paper we propose a new community-based collaborative filtering recommendation algorithm. The main contribution of this paper are following:
\begin{itemize}
  \item A fuzzy community detection is proposed which can find overlapping communities using the PageRank metaphor. 
 \item We propose a novel a recommendation system which utilized community structures to provide recommendation based on proposed fuzzy community detection. In this method, both local (community based) and global similarities between users are considered and the recommendation is based on the community similarities and the rating similarities.

 The results show that the proposed method can outperform some of the latest recommendation systems. 
  
\end{itemize}

The rest of this paper is structured as follows: \autoref{sec:1} is devoted to the reviewing the related works in literature. In \autoref{sec:2}, the proposed method is explained.  \autoref{sec:3} evaluates the new method and compares it with several previous works. Finally, we conclude the paper and present future research directions in \autoref{sec:4}.

\section{Related Work} \label{sec:1}

In this section, we will review recent related work of collaborative filtering and community-based recommender systems. 

Park et al. \cite{park2015reversed} presented a Reversed collaborative filtering (RCF), a fast collaborative filtering algorithm which uses k-nearest neighbor (k-NN) graph. The primary idea of this approach is that it reverses the procedure of finding $k$neighbors. As a result, they provide a novel approach that has a reduction in preprocessing time and recommendation time compared to traditional user-based and item-based CF algorithms. 

Choi et al. \cite{choi2013new} suggested a new similarity function to improve the recommendation quality with the main idea of selecting different neighbors of a target user for each various target item. They implement this idea in two general steps: first, they calculated the item similarity between the target item and each of the co-rated items, and then used it as a weight of each co-rated item while calculating the user similarity between the target user and every other user. Their experiments showed that their approach results in more accurate recommendation than traditional CF models.

Liu et al. \cite{liu2014new} proposed a new user similarity model in order to improve recommendation performance when ratings are not enough to calculate the similarities for each user. In other words, their heuristic similarity measure model is improved, because it captures the effective similar users especially for a cold user who rates a few numbers of items. Their novel similarity measure approach is based on three factors of similarity, the Proximity, the Impact and the Popularity, so the measure is named PIP. It takes the proportion of the common rating between two users into account. Remarking different users have various rating preferences, the proposed measure uses the average and variance of the rating to determine the rating preference of the user. The experimental results demonstrated that the new similarity measure outperforms most other methods and its effectiveness can overcome the disadvantages of the traditional similarity measures.

Kaleli \cite{kaleli2014entropy} introduced a new entropy-based neighbor selection approach for collaborative filtering. They consider that it is very important to use all possible information in order to improve prediction quality, because user-item matrix in recommender systems are very sparse. Also, they incorporated degree of uncertainty (DU) and similarity information to make more proper nearest neighbor (NN) of an entity. In their proposed method, they aim to form a neighborhood which must have maximum similar entities with minimum uncertainty difference. As a result, the new neighborhood selection approach faces with an optimization problem similar to 0-1 knapsack problem which is one of the most popular combinatorial optimization problem. They evaluate method's accuracy in comparison with traditional CF algorithms by performing benchmark data sets-based experiments. The experimental results showed that their method can improve recommendation accuracy of CF algorithms and it is possible to integrate it with other compatible prior methods. 

Zhang et al. \cite{zhang2014collaborative} proposed a framework in which they built the user preference models based on item domain features and then combine these models with collaborative filtering for personalized recommendations, and also the framework helps to discover implicit relationships among users which are missed in traditional CF methods. This framework includes three algorithms: the UPM-B-IDF (modeling user preferences matrix based on item domain features) which intends to use item domain features to model user preferences matrix; the UPV (modeling user preferences vector) which means to formative the user preferences vector from user preferences matrix; and CF-B-UCM (CF based on the user preference model) that aims to incorporate user preference models with CF to make personalized recommendations. Their empirical results showed that the method gets better results and prove the user preference model is more effective for the recommendation.

Ghavipour and Meybodi \cite{ghavipour2016adaptive} has introduced a fuzzy method in which, the level of users trust to each other is defined as fuzzy. For this purpose, they propose a method to adjust membership functions of fuzzy trust and distrust in recommender systems by using learning automata.

Ghazanfar et al. \cite{ghazanfar2014leveraging} presented a clustering algorithm to solve gray-sheep users problem in recommender systems. They demonstrated that collaborative filtering algorithms fail to make accurate recommendations for gray-sheep users, so they proposed k-means clustering algorithm to identify these users and make reliable recommendations for them by using their content-based profiles. They also introduced new improved centroid selection approaches and distance measures for the k-means clustering algorithm. The results showed that the centroid selection approaches did not considerably affect the cluster quality; but the distance measure can alter the performance of the clustering algorithm.

Liu et al. \cite{liu2016cogrec} proposed a Community-Oriented Group Recommendation framework (CoGrec) to make recommendation to group users. In the proposed framework, they utilize non-negative matrix factorization in order to discover overlapping communities, and also they offer four aggregation strategies along with some allocation strategies to generate better recommendations. As a result, the extensive experiments showed the effectiveness of the proposed framework and additionally, it had a better interpretation than most other automatically grouping methods that apply the interest-based clustering approach.

Ding et al. \cite{ding2015community} introduced a novel community detection algorithm based on topological potential theory, which places the users with similar tastes into the same community and then incorporates it with traditional collaborative filtering algorithm. The proposed approach, search for nearest neighbor in the same community with the target user instead of the whole network which limits the search and improves the prediction accuracy. The empirical experiments demonstrated that this method decreases the effect on the prediction accuracy of the sparse matrix, and improves the prediction ability of a recommender system very much.

Cen Cao et al. \cite{cao2015improved} presented an improved collaborative filtering based on community detection (CFCD). Firstly, they mapped user-item matrix into the user similarity network. Then, they adopted a novel discrete particle swarm algorithm (NPSO) to find communities in the user similarity network, and finally they recommend Top-N items to the recommended user according to the communities. The proposed method reduces the amount of computation in neighbor selection, and also it can achieve results without prior knowledge of the size and number of communities. As a result, this algorithm has a great performance in precision, coverage and efficiency of recommendation.

Fulan Qian et al. \cite{qian2013community} proposed a new Community-based User domain Collaborative Recommendation Algorithm (CUCRA). This algorithm is performed in two section: firstly, it builds the offline user domain model; secondly, it recommends items to target users in the model by applying collaborative filtering. The former section consists of three steps: (1) calculate user similarities using a user-item preference dataset; (2) transform a user-item dataset into user-user social networks with the KNN method; (3) find communities with similar user preferences to define a user domain model using community detection methods. This method has a perfect online performance since it recommends items to users in communities instead of to whole social network. Results showed that the time-complexity of the algorithm was reduced to \textit{O (n}).

Guo \cite{guo2013integrating} proposed three various approaches from the point of view of preference modeling to alleviate  data sparsity and cold start problem. Low accuracy and coverage also are the issues of recommender system which have insufficient ratings. So this work addresses these issues in his proposed method, too. Firstly, it combines the ratings of trusted neighbors and makes a new rating profile for the active users. Secondly, it introduces a new Bayesian similarity measure in order to make better use of user ratings. Thirdly, it eliminates the concerned issues by proposing a new information source based on virtual product experience in virtual reality environments.

Guo et al.\cite{guo2015leveraging} proposed a multi-view clustering method to address the low accuracy and coverage in clustering-based recommender systems. In this method, users are iteratively clustered from the views of both rating patterns and social trust relationships.

Alizade and Sheugh \cite{sheugh2015merging} proposed a multi-view clustering based on Euclidean distance by combining similarity-based distances and trust-based distances. This method reduces low accuracy and coverage in cluster-based recommender systems.

\section{Fuzzy Community-Based Recommender System}\label{sec:2}

In this paper, we propose a community-based collaborative filtering which use local and global similarities. Our proposed method consists of three parts: (1) creating the  network of users, (2) community detection based on personalized PageRank, and (3) recommending items to users based on extracted community information. The community detection method proposed in this paper uses fuzzy community detection concepts. In traditional community detection algorithms, system assigns each user to a single group and causes a unique membership for each user, however the fuzzy community detection algorithm provides the membership value or belonging degree for each user. This membership value is used in the recommendation phase to provide a better recommendation and prevent system from losing concepts thanks to high flexibility of proportional membership.  

\subsection{Creating The  Network of Users}

Our recommender system is based on the community detection. Thus, in this approach we generate a network that reveals the relation between the users and recognize their communities. In order to generate a network between users, we consider the user-item rating matrix $R$. The user-item matrix in the recommender systems contains the ratings of items given by users. The user-item rating matrix $R$ for a recommender system with \textit{m} users and \textit{n} items can be considered as follows:

\begin{equation} 
R= \left[ \begin{array}{cccc}
r_{11} & r_{12} & \dots  & r_{1n} \\ 
r_{21} & r_{22} & \dots  & r_{2n} \\ 
\dots  & \dots  & \dots  & \dots  \\ 
r_{m1} & r_{m2} & \dots  & r_{mn} \end{array}
\right]
\end{equation} 

where $r_{ij}$ is the rating of user $i$ on item $j$. We assume that the rating score is greater than zero and if the user $i$ does not rate on item $j$, the value of $r_{ij}$ is 0. 

A user similarity matrix can be generated from the user-item rating matrix $R$. The user similarity matrix expresses the similarities between users. The user similarity matrix $S$ is defined as follows:

\begin{equation}
S_{U = }\left[ \begin{array}{cccc}
s_{11} & s_{12} & \dots  & s_{1n} \\ 
s_{21} & s_{22} & \dots  & s_{2n} \\ 
\dots  & \dots  & \dots  & \dots  \\ 
s_{n1} & s_{n2} & \dots  & s_{nn} \end{array}
\right]
\end{equation}

where $s_{ij}$ is the similarity between user $i$ and user $j$ and defined as the number of common items rated by user $i$ and user $j$.

 Finally, we construct the user network by generating its adjacency matrix. The adjacency matrix  of the user network is defined as follows:
\begin{equation}
A= \left[ \begin{array}{cccc}
a_{11} & a_{12} & \dots  & a_{1n} \\ 
a_{21} & a_{22} & \dots  & a_{2n} \\ 
\cdots  & \dots  & \dots  & \dots  \\ 
a_{n1} & a_{n2} & \dots  & a_{nn} \end{array}
\right]
\end{equation}

 where $a_{ij}$ is assigned based on a similarity threshold $\tau$. \textit{a${}_{ij}$=1} if \textit{s${}_{ij}$} is larger than $\tau$, and \textit{a${}_{ij}$=0} otherwise. Thus, if the element $A_{ij}$ of matrix $A$ is 1, it means there is a link between the user $i$ and user $j$, but if it is 0 means there is no connection between them.

\subsection{Novel Fuzzy Community Detection Method}

In the proposed method, the community detection is utilized to find similar users.
In order to detect the community of users, we propose a new fuzzy community detection method. 
In this method, community is defined as a group of nodes in a complex network which are strongly connected to each other. The nodes belonging to the same community are more similar and thus users in the same communities probably have similar tastes and interests. Hence, by applying community detection techniques, we may produce better recommendation results. 

Considering the above facts, we tried to implement a fuzzy community detection system using personalized PageRank. We utilize the fuzzy results of the community detection for the recommendation task.  

The proposed community detection method implemented in two phases: the first phase calculates personalized PageRank for all nodes in the user network, and the second phase clusters the nodes (users) based on the PageRank values by the fuzzy c-mean method. 

\subsubsection{PageRank}

The PageRank algorithm is a method that assigns a real number to each page (node) of the web (graph) based on the network structure of the web, in order to measure the relative importance of the web pages (nodes) \cite{page1999pagerank}. The PageRank algorithm expresses that if a page has important links to it, its links to other pages also become important. Therefore, a page receives an amount of rank from every page which links to it and propagates the rank through its outgoing links. In this sense, a page has a high rank if it is referenced by high rank pages \cite{haveliwala2003topic}.

 The basic idea of PageRank is that if page $u$ has a link to page $v$, then page $u$ gives some contributions to page $v$. Let $u$ be a web page, $B(u)$  be the set of pages pointing to $u$. Moreover, let $R(u)$ and \textit{$R(v)$ } be the rank scores of the page \textit{$u$ }and \textit{$v$ }, respectively. Let $N_{v}$ be the degree of the page $v$. Therefore, the simplified PageRank equation for $u$ is recursively defined as:
\begin{equation} \label{GrindEQ__1_} 
R(u)=\sum _{v\in B(u)}^{}\frac{R(v)}{N_{v} }   
\end{equation} 

By considering the above equation for each node $u$ of a network, we have a system of equations. This system can be written in a Matrix format as follows:

\begin{equation} \label{GrindEQ__100_} 
R=M R  
\end{equation}
   By adding another equation stating that the sum of the ranks is one $\sum _{v\in B(u)}=1$, it is possible to solve the system of the equations by the Gaussian elimination method \cite{page1999pagerank}.  However, this method is not efficient for a large network. However, the power iteration method provides an efficient  way to find the PageRank of each node.

The equation \autoref{GrindEQ__100_} may be computed by staring with any set of ranks and iterating the computation until it converges.
where $M$ is the transition matrix which is derived from the adjacency matrix of the network, and the vector $R$ us the PageRank vector which contains the PageRank of each node. The \textit{jth} component in \textit{R} demonstrate the PageRank value of page \textit{j}. If the transition matrix is a column stochastic matrix then the dominant eigenvector of $M$ is one and thus the vector $R$ is the dominant eigenvector of $M$. 
The matrix $M$ is a square, stochastic matrix with the rows and columns corresponding to the directed graph (network) $G$ of the web, supposing all nodes in $G$ have at least one outgoing edge. The matrix element $m_{ij}$ in row $i$ and column $j$ has value $\frac{1}{D} $ if page $j$ has $D$ arcs out, and one of them is to page $j$ otherwise, $m_{ij} =0$. Computation of the equation \ref{GrindEQ__1_} corresponds to the matrix calculation: \textit{R= MR}. As the vector \textit{R} is the dominant (principal) eigenvector of the matrix $M$, it is possible to find it by the power iteration method. \textit{R} can be calculated by applying M to an initial vector  $R=[{{1}/{N]}}_{N\times 1}$ frequently. In fact, repeatedly multiplying \textit{R} by \textit{M} yields the dominant eigenvector of matrix $M$. If the matrix $M$ is not column stochastic i.e. there is a node in the graph with no outgoing link (this node is called dead end), after each iteration of the power iteration method the sum of the PageRank values becomes less than one. Thus if the power iteration continues in this way the methods converge to a PageRank vector that the ranks of all nodes are zero. In order to fixed this problem if the network has dead end after each iteration of the power iteration method the amount of the rank sunken is added to all nodes. In other words, after each iteration of the power iteration the following procedure is needed. 

Until here we used tricks to simply PageRank algorithm as much as possible there is still a problem with the simplified PageRank. Consider a group of pages that all point to each other but to no other page, and suppose there are some pages that point to one of them. Then, during the iteration, this loop would accumulate PageRank values but never distribute any PageRank values. This scenario is called a rank sink. 

To solve the rank sink problem, another matrix $M'$ was defined, which transition edges of probability$\ $$\frac{{\rm \beta }}{N} $ between every pair of nodes in $G$ are added to $M$. The equation \ref{GrindEQ__2_} shows a new matrix $M'$.
\begin{equation} \label{GrindEQ__2_} 
M'={\rm \beta }M+(1-{\rm \beta )}\left[\frac{1}{N} \right]_{N{\rm \times }N}  
\end{equation} 
Where ${\rm \beta }$ $\ $is a  dampening factor that is usually in the range 0.8 to 0.9. This modification improves the quality of PageRank by introducing a factor ${\rm 1-\beta }$ which limits the effect of rank sinks. 

%Computing PageRank values of all pages is defined in equation \ref{GrindEQ__3_}.
%%%\end{equation} 

\subsubsection{Personalized PageRank}

There are several improvements could be made to PageRank. Personalize PageRank is a variant of the original PageRank algorithm.  Personalized PageRank models the relevance of nodes in a network from the point of view of a given node. It has applications in search engine, community detection, and other applications.  

The personalized PageRank of each node based on the node  $i$ is calculated as follows:

\begin{equation} \label{GrindEQ__4_} 
R={\rm \beta }M{\rm}R+(1-{\rm \beta })V 
\end{equation}
where $V$ is a vector with $n$ elements and only $V_i$ is equal to 1 and other elements are zero.
 By this method, a part of the rank is given to the node $i$ and this rank is propagated to the  nodes which are nearer to the node $i$ and have more connection to it. Thus the ranks of the nodes which are more related to the node $i$ become greater than their standard PageRanks.

\subsubsection{Detecting Communities}
In order to detect the communities (clusters) in the user network, the Pagerank-based fuzzy community detection considers several features for each user. All users in the user network can be described by $n$ features (dimensions) each of which is the personalized PageRank of that user with respect to one of the other users. In other words user $i$ can be described by the vector $PR_i=\{pr_{i1},pr_{i2},\dots,pr_{in}\}$   where $pr_{ij}$ denotes the personalized PageRank of user $i$ with respect to the user $j$. Thus, we have a dataset with $n$ instances (vectors) and $n$ dimensions and the fuzzy c-means method is used to cluster this dataset.  

As this data is high dimensional, before the clustering, a pre-processing job must be performed on raw data to achieve an acceptable set of input data. Therefore, the principal component analysis (PCA) is applied in order to reduce the dimensions and also adjust any unwanted correlation in the dataset. After the pre-processing phase, the fuzzy c-means algorithm is applied on the dataset in order to build communities of the users. The intuition behind this approach is that if two users have similar personalized PageRank vector, they are likely to be in the same cluster.

Fuzzy c-means clustering (also called soft clustering) is a kind of clustering that may assign a data point to more than one cluster. Clustering is the process of assigning data points to homogeneous clusters, as items in the homogeneous cluster are similar to each other as much as possible. Clusters are identified by similarity measures. Different similarity measure may be selected  on the basis of data or usage. In crisp clustering (which is also known as hard clustering) dataset is partitioned into distinct clusters, in a way that each data point exactly belongs to only one cluster. However, in the fuzzy clustering, data points potentially can belong to several clusters. One of the most popular fuzzy clustering algorithms is fuzzy c-means \cite{bezdek1984fcm}.

This algorithm performs on a restricted set of $n$ elements $X=\left\{x_{1,....,} x_{n} \right\}$ and divides the whole network into $c$ fuzzy clusters, according to a certain scale. On a limited specific set, this algorithm returns a list of $c$ cluster centers $C=\left\{c_{1} ,...,c_{c} \right\}$ which are weighted average of member points, center of the cluster $k$ is obtained from equation \ref{GrindEQ__5_}, where \textit{$w_{k} (x)$} represents the belonging degree of point \textit{x} to cluster \textit{k} and m determines the level of cluster fuzziness.

\begin{equation} \label{GrindEQ__5_} 
C_{k} =\frac{\sum _{x}^{}w_{k} (x)^{m} x }{\sum _{x}^{}w_{k} (x)^{m}  }  
\end{equation} 
Fuzzy c-means algorithm is also returns matrix \textit{$W=w_{ij} \in \left[0,1\right]$} that \textit{$ i=1,2,\ldots ,n$} and \textit{$j=1,2,\ldots,c$}. Each element of $w_{ij}$ represents the level of membership degree of the element $X_{i}$ to cluster $c_{j}$. In other words, this algorithm, assigns a membership degree to each data point for each centroid, which is based on the distance between centroid and the data point. The more data is close to a centroid, belonging degree to that cluster is more. Total belonging degree of each data point must equal to 1, and it is calculated from equation \ref{GrindEQ__6_}:
\begin{equation} \label{GrindEQ__6_} 
w_{ij} =\frac{1}{\sum _{k=1}^{c}\left(\frac{\left\| x_{i} -c_{j} \right\| }{\left\| x_{i} -c_{k} \right\| } \right)^{\frac{2}{m-1} }  }  
\end{equation} 
The principal purpose of the fuzzy c-means algorithm is to minimize the objective function mentioned in equation \ref{GrindEQ__7_}:
\begin{equation}\label{GrindEQ__7_}
{argmin_{W}} \sum _{i=1}^{n}\sum _{j=1}^{c}w_{ij}^{m}   \left\| x_{i} -c_{j} \right\| ^{2}
\end{equation}

Fuzzy clustering pseudo code can be seen in algorithm \autoref{alg1}. By considering \textit{$X=\left\{x_{1}, \ldots x_{n} \right\}$ }as data points set and $C=\left\{c_{1} ,\ldots,c_{c} \right\}$ as cluster centers set.

% 1) cluster center c randomly is selected.

% 2) belonging degree ${}_{w_{ij} }$${}_{ }$is calculated from equation \eqref{GrindEQ__6_}

% 3) cluster center$c_{k} $ ${}_{  }$is calculated from equation \eqref{GrindEQ__5_}

% 4) repeat steps 2 and 3 until the equation \eqref{GrindEQ__7_} is minimized.

\begin{algorithm}
 \begin{algorithmic}

\STATE\textbf{Input:} Users Set

\STATE\textbf{Output:} Membership of users to each of the classes
 
\STATE \textbf{Step 1:} Centroids are initiated randomly
\STATE \textbf{Step 2:} Belonging degree $w_{ij}$ is calculated from equation \ref{GrindEQ__6_}
\STATE \textbf{Step 3:} Centroids $c_{k}$ is calculated from equation \ref{GrindEQ__5_}
\STATE \textbf{Step 4:} Repeat steps 2 and 3 until the equation \ref{GrindEQ__7_} is minimized.

 \end{algorithmic}
 \caption {Membership calculation of users for each classes}\label{alg1}
\end{algorithm}

\subsection{Creating Recommendation Criteria}

After building the communities, the method predicts the ratings and recommends items for users on the basis of combination between community detection and collaborative filtering (considering the similarity between user scoring). The basic idea of user-based collaborative filtering is using the similarity among users for the recommendation. In the proposed approach, we calculated similarities between users with two methods. At first, users belonging to the same community considered as a set of users with similar preferences. As the fuzzy c-mean method is used for community detection, the belonging degree of two users to the same cluster may be different. Thus, to compute the community similarity (local similarity) between two users the sum of multiplication of the belonging degrees of each two users to different clusters is considered.

As a global similarity measure, the method uses the correlation similarity measure to compute the similarity between users. The correlation similarity measure between every two users is defined as equation \ref{GrindEQ__8_}  

\begin{equation} \label{GrindEQ__8_} 
COR(u_{a} ,u_{b} )=\frac{\sum _{i\in {\rm I} _{u_{a} ,u_{b} } }^{}(r_{u_{a},i } -\bar{r}_{u_{a} } )(r_{u_{b} ,i} -\bar{r}_{u_{b} } ) }{\sqrt{\sum _{i\in {\rm I} _{u_{a} ,u_{b} } }^{}(r_{u_{a},i } -\bar{r}_{u_{a} } )^{2} \sum _{i\in {\rm I} _{u_{a} ,u_{b} } }^{}(r_{u_{b} ,i} -\bar{r}_{u_{b} } )^{2}   } }  
\end{equation} 

where $i\in {\rm I} _{u_{a} ,u_{b} } $  represents the set of all items that were rated by both users $u_{a} $ and $u_{b} $. $\bar{r}_{u_{a}}$ is the average of the rating that user $u_a$ has given to the items and $r_{u_{a},i}$ is the rating of the item $i$ given by user $u_a$.

Note that all the users in the same community do not have the same weight in rate prediction, in correlation measure the rating of the nearest user to the target user has more weight and vice versa. Moreover, if the belonging degree of a node to a cluster was more, it has a more weight in rate prediction. Pearson correlation between two users $u_{a} $ and $u_{b} $ is calculated as equation \ref{GrindEQ__8_}.

Then in order to predict the rank that user $u_i$ will give to the item $j$ equation \ref{GrindEQ__9_} is  used.

\begin{equation} \label{GrindEQ__9_} 
rank(u_{i} ,{j} )=\frac{\sum _{k\in n}^{}\varpi(u_{i},u_k) \times rank(u_{k} ,i_{j} ) }{\sum _{k\in n}^{}\varpi(u_{i},u_k) }  
\end{equation}

Where w${}_{k}$${}_{  }$is calculated from equation \ref{GrindEQ__10_}:
\begin{equation} \label{GrindEQ__10_} 
\varpi(u_{i},u_k) =\alpha \sum _{c_l \in c_{\bigcap } }^{n}\mu _{u_i,c_l}.\mu _{u_k,c_l} +\beta.COR(u_i,u_k) 
\end{equation} 
Here $\alpha $ and $\beta $ are the factors that determine the impact of each measures on the weight of  Rank, then optimized value of these factor will be obtained in experiments. $\mu _{u_i,c_l} $ and $\mu _{u_k,c_l} $  represent the belonging level of user ${u_i}$ and user $u_k$ to the cluster $c_l$ respectively, additionally $c_{\cap }$ shows the clusters which the belonging level of both user I and k to those clusters are more than a threshold $\theta$. in other words, $c_{\cap }$ is a set of clusters that  the equation \ref{GrindEQ__11_} is true for them:
\begin{equation} \label{GrindEQ__11_} 
\text{if    } ( \mu _{u_{i} ,c_l} {\rm \& }\mu _{u_{k} ,c_l} {\rm >}\theta)  \text{   } \to c_l  \in c_{\bigcap }  
\end{equation} 

Pseudo-code of the proposed method is shown in algorithm \autoref{alg2}.

\begin{algorithm}
 \begin{algorithmic}

\STATE\textbf{Input:} Users' ratings on items, records to be predicted

\STATE\textbf{Output:} Results of the prediction
 
\STATE \textbf{Step 1:} Create user-item rating matrix  $R=\left[ \begin{array}{cccc}
r_{11} & r_{12} & \dots  & r_{1n} \\ 
r_{21} & r_{22} & \dots  & r_{2n} \\ 
\dots  & \dots  & \dots  & \dots  \\ 
r_{m1} & r_{m2} & \dots  & r_{mn} \end{array}
\right]$  on the basis of users' ratings on items. 

\STATE \textbf{Step 2:} Map a user-item network \textit{R} into user similarity network S${}_{U = }$$\left[ \begin{array}{cccc}
s_{11} & s_{12} & \dots  & s_{1n} \\ 
s_{21} & s_{22} & \dots  & s_{2n} \\ 
\dots  & \dots  & \dots  & \dots  \\ 
s_{n1} & s_{n2} & \dots  & s_{nn} \end{array}
\right]$ and build a user-user relation matrix. 

\STATE \textbf{Step 3:} Set a similarity threshold $\tau$ and construct the user network adjacency matrix  $A=\left[ \begin{array}{cccc}
a_{11} & a_{12} & \dots  & a_{1n} \\ 
a_{21} & a_{22} & \dots  & a_{2n} \\ 
\cdots  & \dots  & \dots  & \dots  \\ 
a_{n1} & a_{n2} & \dots  & a_{nn} \end{array}
\right]$.

\STATE \textbf{Step 4:} Computing the personalized PageRank matrix ($PR$) of the user network ($A$).

\STATE \textbf{Step 5:} Use PCA to reduce the dimensions of $PR$ matrix.

\STATE \textbf{Step 6:} Apply fuzzy C-means clustering algorithm on $PR$ after dimensionality reduction to build the community, and put users with similar personalized PageRank into the same community.

\STATE \textbf{Step 7:} Use correlation measure $COR(u_{a} ,u_{b} )$ in each community detected in step 7 to apply weighted mean  $w_{k} $ for predict rating to the target user.

\STATE \textbf{Step 8:} Compute the predicted rating  for user on the basis of step 7.

 \end{algorithmic}
 \caption {Fuzzy Community-based Collaborative Filtering Algorithm Description}\label{alg2}
\end{algorithm}

\newpage
\section{Experimental Results}\label{sec:3}

On the following, we analyze recommendation that has been made by our new approach and compare it to several recent approaches that also achieved high performance in their recommendation tasks.

In this paper, three categories of experiments on the MovieLens  and FilTrust datasets has been performed to evaluate proposed algorithm performance.

In all the experiments, a fixed value for similarity threshold is defined and addressed as $\tau$. This threshold will be used as a controller for establishing communication between users in the network. The threshold $\tau$ is chosen in such a way that the clustering approach has the appropriate modularity. According to initial experiments, a suitable threshold for MovieLens dataset is 15 and for Filmtrust it will be 7. 

In other words, to create a communication network, adjacency matrix $A$
% $A = \left[ \begin{array}{cccc}
% a_{11} & a_{12} & \dots  & a_{1n} \\ 
% a_{21} & a_{22} & \dots  & a_{2n} \\ 
% \cdots  & \dots  & \dots  & \dots  \\ 
% a_{n1} & a_{n2} & \dots  & a_{nn} \end{array}
% \right] $ 
is calculated  from  user-user relation matrix $S_{U} = \left[ \begin{array}{cccc}
s_{11} & s_{12} & \dots  & s_{1n} \\ 
s_{21} & s_{22} & \dots  & s_{2n} \\ 
\dots  & \dots  & \dots  & \dots  \\ 
s_{n1} & s_{n2} & \dots  & s_{nn} \end{array}
\right] $ where each element in $A$ is determined based on the similarity threshold value. If \textit{$s_{ij}$} is larger than threshold $\theta $, $a_{ij}$ is 1, otherwise $a_{ij}$ is equal to 0. Value 1 means that there is a link between user $i$ and user$j$ , value 0 means that there is no connection between user $i$ and user$j$.

\subsection{Datasets}

The experiments are carried out on MovieLens and FilmTrust datasets to verify recommendation results. Both mentioned datasets are among well-known datasets with no unusual behavior in samples recorded also, due to the fact that proposed method must be compared to other approaches, using common datasets with other papers will make comparison task more reliable.

The MovieLens dataset was collected by the GroupLens research project at the University of Minnesota. In ML-100K dataset, there are 100,000 ratings with 943 users and 1682 movies and the score of ratings are on a scale of 1 to 5.  Each user has rated at least 20 movies. Also, this dataset consist of simple demographic info for users includes age, gender, occupation, and zip \cite{zhou2015social}. However, in our experiments, we only use user id, item id, and rating similar to most of the previous works. For evaluating the performance of the proposed recommender system, the dataset is divided into two parts, 80\% as the training data and the remaining  20\% as the test data.

The FilmTrust dataset  is a social website based on trust that users can rate films and criticize them. Due to user privacy conservation, data set is not publicly available, the whole website is crawled in June 2011 and the number of 1508 users, 2071 items and 35497 ratings have been obtained. The ratings values are 0.5, 1, 1.5{\dots}.3.5, and 4 \cite{ghazanfar2010scalable}. Similar to MovieLens, this dataset is also divided into two parts 80\% and 20\%, for training and test respectively. So the train-evaluation ratio is the same for both datasets.

\subsection{Evaluation Metric}

In order to evaluate the accuracy of the proposed approach, two popular metrics are used: \textit{Mean Absolute Error} (MAE) and \textit{Root Mean Square Error} (RMSE). MAE measures the prediction accuracy by calculating the deviation between the experimental results and real data. A smaller MAE and RMSE means recommender system predicts user ratings more accurately

 The metric MAE is defined as equation \ref{GrindEQ__12_}:
\begin{equation} \label{GrindEQ__12_} 
MAE=\frac{\sum _{(u_i,j)\in T}^{}\left|rank_{(u_i,j)} -\overline{rank}_{(u_i,j)} \right| }{\left|T\right|}  
\end{equation}

Where ${T}$ denotes the test data, and $rank_{(u_i,j)} $$\ $ is the real rating of user ${u_i}$ on item ${j}$, while $\overline{rank}_{(u_i,j)} $  is the predicted rating of the system.

\subsection{Clusters Quantity Impact Analysis}
Here, the accuracy of the method is investigated with the different number of clusters.  Fuzzy c-means is sensitive to the number of clusters, so taking mentioned fact into the account, the efficient number of clusters must be achieved in an iterative process. Therefore, this experiment investigates the impact of the number of clusters in our recommender system. 

\begin{figure}[!ht]
\centerline{{\includegraphics[scale=0.6]{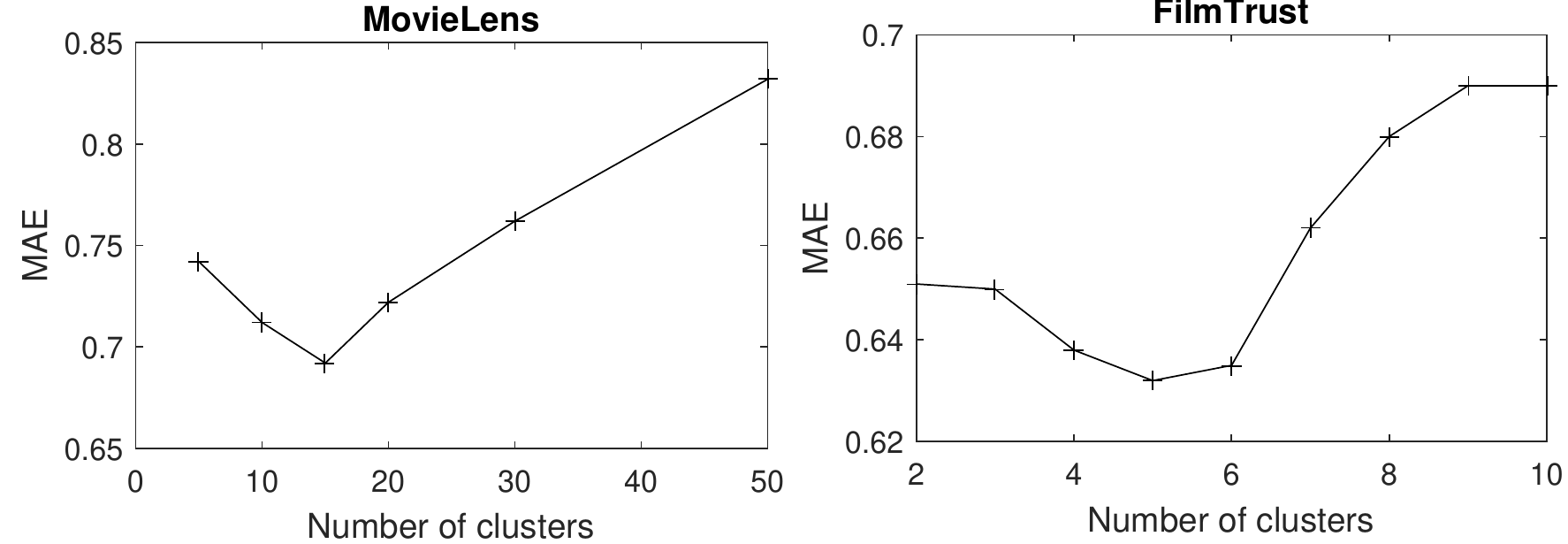}}}
 \caption {The effect of the number of clusters (communities)}
\label{fig:k}
\end{figure}

\autoref{fig:k} shows different values of MAE for different clusters of MovieLens  dataset.

In this approach, when the number of clusters is few, poorer outcomes are achieved because there is a chance that people who are not so similar together exist in the same cluster. On the other hand, when the number of clusters increases so much, the number of people who are in the same cluster is reduced and the system cannot generalize recommendation for desired user respect to its society within the cluster and consequently, the error rate will increase. As it is shown in \autoref{fig:k}, for 15 number of clusters, the most optimal MAE is achieved.

Moreover, \autoref{fig:k} also shows different values of MAE for different clusters of FilmTrust dataset. In \autoref{fig:k} for 5 number of clusters, the most optimal MAE is gained. The same discussed phenomena that happened for MovieLens dataset can be seen here, when the number of clusters are too many or too few, MAE may increase, and for obtaining lowest MAE optimal value of the cluster numbers should be found.

\subsection{Coefficients Combination Impact Analysis}

To predict user $i$'s rating for item $j$, a weighted mean was used which calculated by equation \ref{GrindEQ__10_}. So the accuracy level of the proposed algorithm can be different for various values for $\alpha$ and $\beta$. Therefore, in this experiment, we study the impact of these coefficients. For simplicity we define  $\gamma =\frac{\alpha }{\beta } $ and reports the MAE of the method for different values of $\gamma$. In the previous experiment, the  $\gamma $ was considered as a fixed value (0.3 for MovieLens  and 1 for FilmTrust). In this experiment, we consider the number of cluster as a fixed values, 15 for MovieLens  and 5 for FilmTrust datasets ( the optimal values of previous experiment).

\begin{figure}[!ht]
\centerline{{\includegraphics[scale=0.6]{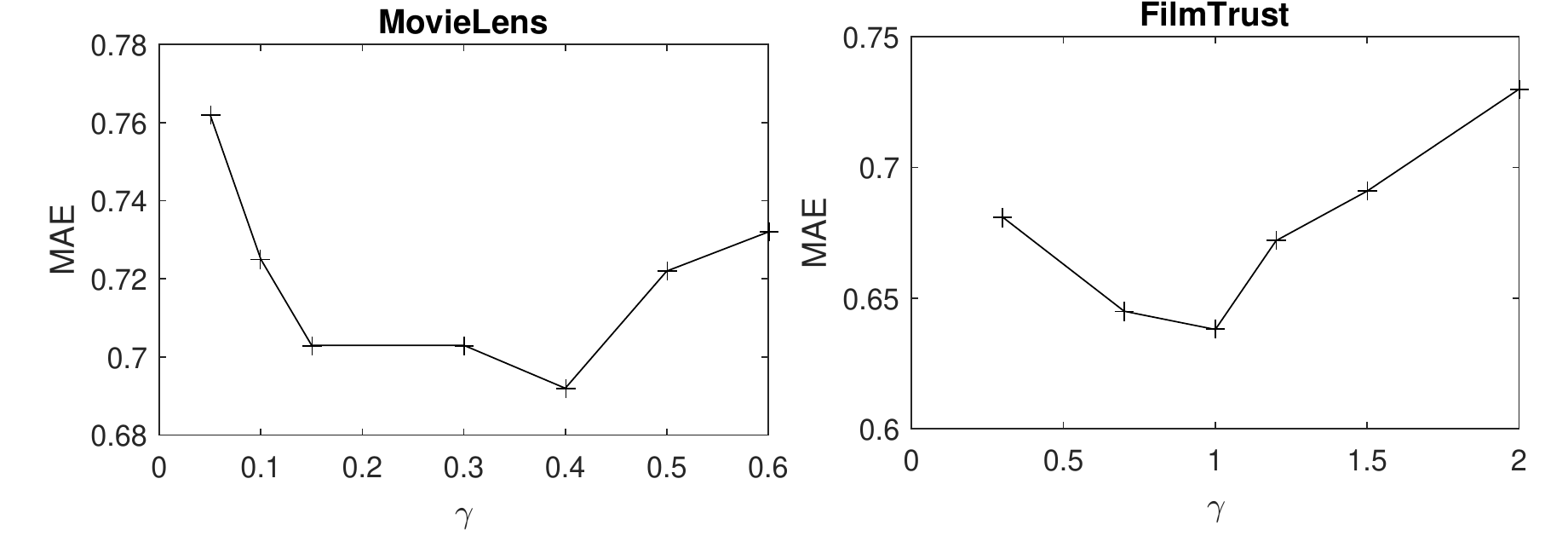}}}
 \caption {The effect of $\gamma$}
\label{fig:gamma}
\end{figure}

\autoref{fig:gamma} shows MAE for different $\gamma $ of  MovieLens  dataset.
In this diagram, it can be seen that for the $\gamma =0.4$ , the most optimal MAE $ =0.692$ is achieved. When the value of  $\gamma $ is reduced, it means that the recommender system considers the correlation similarity more. So when $\gamma $ $\mathrm{\ }$ approaches zero, it means that the recommender system uses only the correlation similarity and when the $\gamma $ get increased, it means that the fuzzy clustering has a more impact in the result of the recommender system. As it is obvious in the diagram, there is a factor that considers the optimal combination of correlation similarity and belonging degree of fuzzy clusters.

Additionally, \autoref{fig:gamma} shows the impact of different $\gamma$ for the FilmTrust dataset.
For $\gamma =1$ the most optimal MAE $ =0.638$ is gained. Also, The same pattern for MovieLens dataset can be seen too. So our thumb rule says that by reduction value of $\gamma $ , the recommender system consider the correlation similarity more, and by increasing this value, fuzzy clustering has more impact on the results.

\subsection{Simulation Results}

In this section, we compare our proposed method with other recent works that have been done on MovieLens and FilmTrust datasets.

In \autoref{tab:11}, the optimal results of our proposed method is compared to other research works evaluated on the MovieLens  dataset.
In \autoref{tab:22} optimal results of our proposed method is compared to other researches on the FilmTrust dataset. Take note that reported numbers are in error format and lower means better also, reported MAE for the proposed model is achieved by running algorithms repeatedly and the average error is reported. The results show that our method is better than most of the previous works.

 \begin{table}[h]
\centering

\caption{MAE of the proposed method in comparison with different recommender systems for MovieLens dataset.}\label{tab:11}
\begin{tabular}{|c|c|}
\hline
Method                             & MAE            \\ \hline
\textbf{Proposed Model}            & \textbf{0.692} \\ \hline
Ghavipour and Meybodi 2016 \cite{ghavipour2016adaptive}  & 0.83           \\ \hline
Ding et al. 2015 \cite{ding2015community}               & 0.845          \\ \hline
Cao et al. 2015 \cite{cao2015improved}                & 0.7206         \\ \hline
Ghazanfar and Prugel-Bennette 2014 \cite{ghazanfar2014leveraging} & 0.758 \\ \hline
Qian et al. 2013 \cite{qian2013community}                   & 0.6919   \\ \hline
\end{tabular}
\end{table}

\begin{table}[h]
\centering

\caption{MAE of the proposed method in comparison with different recommender systems for Film Trust dataset.  }\label{tab:22}
\begin{tabular}{|c|c|}
\hline
Method                   & MAE            \\ \hline
\textbf{Proposed Model}  & \textbf{0.632} \\ \hline
Guo et al. 2015 \cite{guo2015leveraging}          & 0.689          \\ \hline
Alizadeh and Sheugh 2015 \cite{sheugh2015merging} & 0.703          \\ \hline
Guo 2013 \cite{guo2013integrating}                 & 0.605          \\ \hline
\end{tabular}
\end{table}

The recommender systems that compared with our proposed method has been briefly explained in Section 2. 
The results show that our method has a better performance than almost every other proposed models during there years. Performance results for MovieLens dataset that are brought in \autoref{tab:11} show that our proposed algorithm outperforms all the competitive algorithms and has a very tight competition with \cite{qian2013community} in MAE but better time complexity performance since the \cite{qian2013community} suffers from $O(m^3)$ run-time complexity in offline mode also, as you can see in \autoref{tab:22}, Guo \cite{guo2013integrating} achieved better results compared to ours that is obtained by using trusted information as their input information. Due to the fact that many modern digital markets have a huge number of users, it will be a very time-consuming process to check them all and eventually, it will be not possible to perform such a preparation in practical cases. Therefore, we refused to filter the data and tried to use them as natural as they can be. 

\section{Conclusion}\label{sec:4}

In this paper, a novel community detection method was used in order to find similarities among users. For the mentioned task we used Personalized PageRank which caused better detection of communities. In this method,  at first, personalized PageRank is calculated for all nodes of networks, then by using fuzzy c-means algorithm, nodes which have similar PageRanks, are clustered in the same community. After building communities, a weighted mean which uses correlation similarity measure and the belonging degrees to the clusters is utilized in order to predict the ratings. For evaluating the accuracy of the recommendation we used two well-known datasets: MovieLens and FilmTrust. Results show that the hybrid model of recommendation that uses both the community structure analysis and collaborative filtering can give us a better intuition to user relations and interactions and eventually yields better recommendation accuracy. Experiments show that our method successfully outperformed other recent methods and has a better interpreting ability than others in analyzing user-item datasets.

\bibliographystyle{unsrt}
\bibliography{Recommender-Golifroshani_cna4}
\end{document}